\documentclass[journal,twoside,final]{IEEEtran}
\usepackage{graphicx}
\graphicspath{{figures/}}

\RequirePackage{xspace}
\def\PANDA{$\overline{\mbox{P}}${ANDA}\@}

\begin{document}
\bstctlcite{OPTIONS}

\title{Detector developments for the hypernuclear programme at \PANDA}

\author{A.~{Sanchez Lorente},
				P.~Achenbach\IEEEauthorrefmark{1},
        J.~Pochodzalla, and
        S.~{S\'anchez Majos}
\thanks{\IEEEauthorrefmark{1}Presented at the IEEE Nuclear Science Symposium 2008
	(NSS08), Dresden, Germany, 20--24 Oct. 2008.}
\thanks{Manuscript received November 14, 2008. Work supported in part by
  the European Community under the ``Structuring the European Research
  Area'' Specific Programme as Design Study DIRAC\-secondary-Beams
  (contract number 515873).}  
\thanks{The authors are with the Institut f\"ur Kernphysik, Johannes
  Gutenberg-Universit\"at, Mainz, Germany (e-mail: 
  lorente@kph.uni-mainz.de, patrick@kph.uni-mainz.de, 
  pochodza@kph.uni-mainz.de, sanchez@kph.uni-mainz.de)}%
}


\markboth{2008 IEEE Nuclear Science Symposium Conference Record}{N17-7}

\maketitle

\begin{abstract}
  The technical design of the \PANDA\ experiment at the future FAIR
  facility next to GSI is progressing. At the
  proposed anti-proton storage ring the spectroscopy of double
  $\Lambda$ hypernuclei is one of the four main topics which will be
  addressed by the Collaboration. The hypernuclear experiments require (i) a 
  dedicated internal target, (ii) an active secondary target of alternating
  silicon and absorber material layers, (iii) high purity
  germanium (HPGe) detectors, and (iv) a good particle identification system 
  for low momentum kaons. All systems need to operate in the presence of
  a high magnetic field and a large hadronic background. 
  The status of the detector developments for this programme is summarized.
\end{abstract}

\begin{IEEEkeywords}
  Hypernuclei, antiproton-induced reactions, design of experiments.
\end{IEEEkeywords}

\section{The Hypernuclear Programme at \PANDA}
\IEEEPARstart{H}{ypernuclear} research will be one of the main topics
addressed by the \PANDA\ experiment at the planned Facility for
Anti-proton and Ion Research (FAIR) next to the GSI site near Darmstadt.  
The FAIR
complex will include the High Energy Storage Ring (HESR) to store
anti-protons between 0.8 and 14.4\,MeV energy. Intense and high quality
beams with luminosities up to 10$^{32}$\,cm$^{-2}$s$^{-1}$ and
momentum resolutions down to 10$^{-5}$ are expected. The \PANDA\
hypernuclear programme shall reveal the $\Lambda\Lambda$ strong
interaction strength, not feasible with direct scattering
experiments~\cite{Pochodzalla2004,PANDA2005}.

In the planned setup there exist many experimental challenges and
several European research groups are working on the realization of the
detectors.  A detailed design will be available in the mid-term
future. When reflecting upon the state of the preparations, 
one should be aware that the construction of the anti-proton
storage ring and the \PANDA\ experiment has not yet started.

Low momentum $\Xi$ pairs can be produced in $\mathrm{\overline{p}p}
\to \Xi^- \overline{\Xi}^+$ or $\mathrm{\overline{p}n} \to \Xi^-
\overline{\Xi}^0$ reactions with high rates using the anti-proton beam
on an internal target. The advantage as compared to the kaon
induced reaction is the fact that the anti-proton is stable and can be
retained in a storage ring. This allows a rather high luminosity even
with very thin primary targets. The associated $\overline{\Xi}$ will
undergo annihilation inside the residual nucleus.  The annihilation
products contain at least two anti-kaons that can be used as a tag for
the reaction. Due to the large yield of hyperon-antihyperon pairs produced at
\PANDA\ a high production rate of single and double hypernuclei under
unique experimental conditions will be feasible. 

\begin{figure}
  \begin{center}
    \includegraphics[angle=-90,width=\columnwidth]{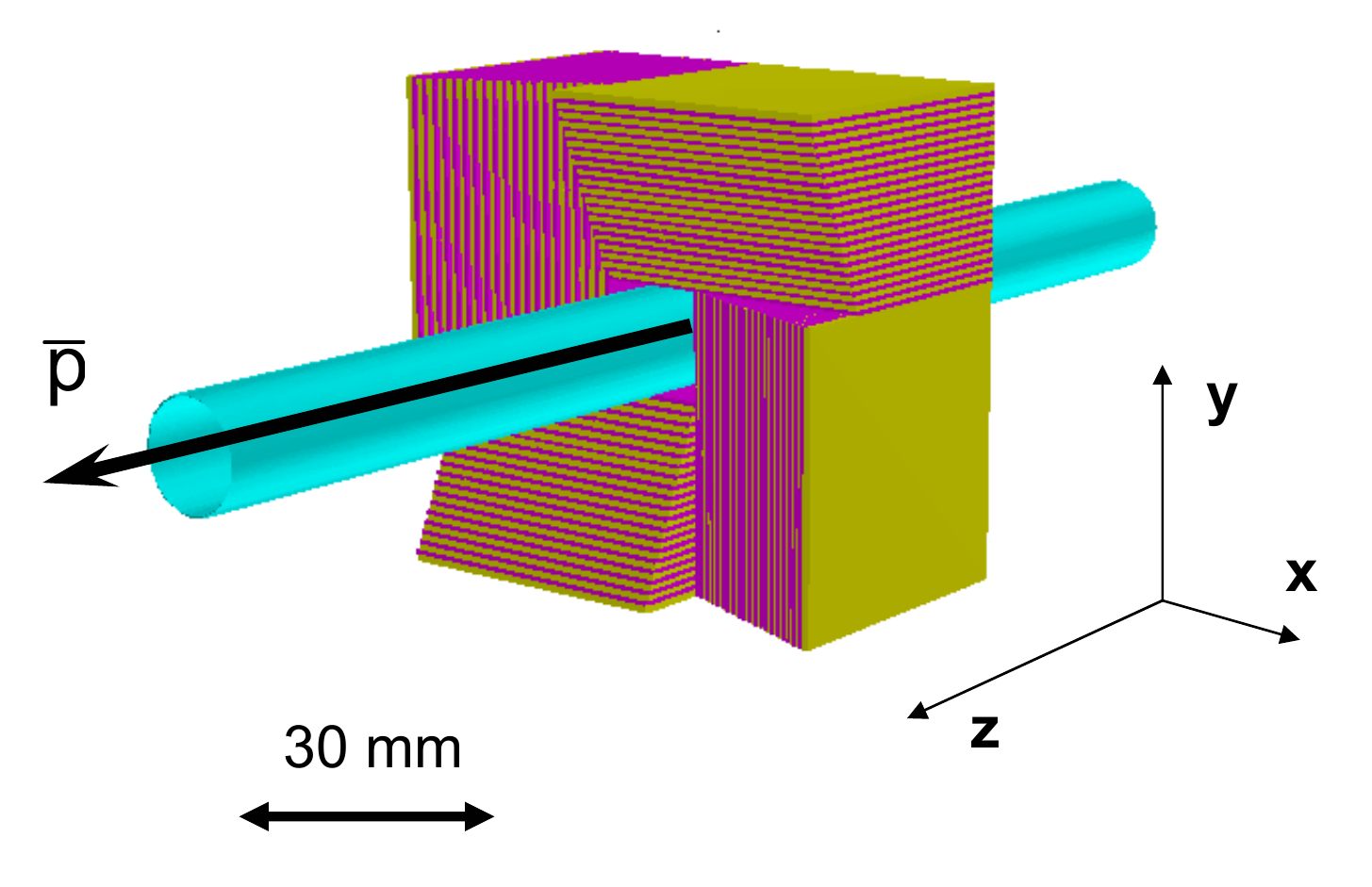}
    \caption{The secondary target consisting of thin layers of silicon
      alternated with different absorber material of light
      nuclei in four segments~\cite{PhDLorente,PhysicsBook}. The target surrounds the
      anti-proton beam-pipe and must not block the trajectories of the
      forward going particles tagging the hypernuclear events (direction
      of anti-protons indicated by arrow).}
    \label{fig:SecondaryTarget}
  \end{center}
\end{figure}

\IEEEpubidadjcol
\section{Hypernuclear Target}
\begin{figure}
  \begin{center}
    \includegraphics[width=\columnwidth]{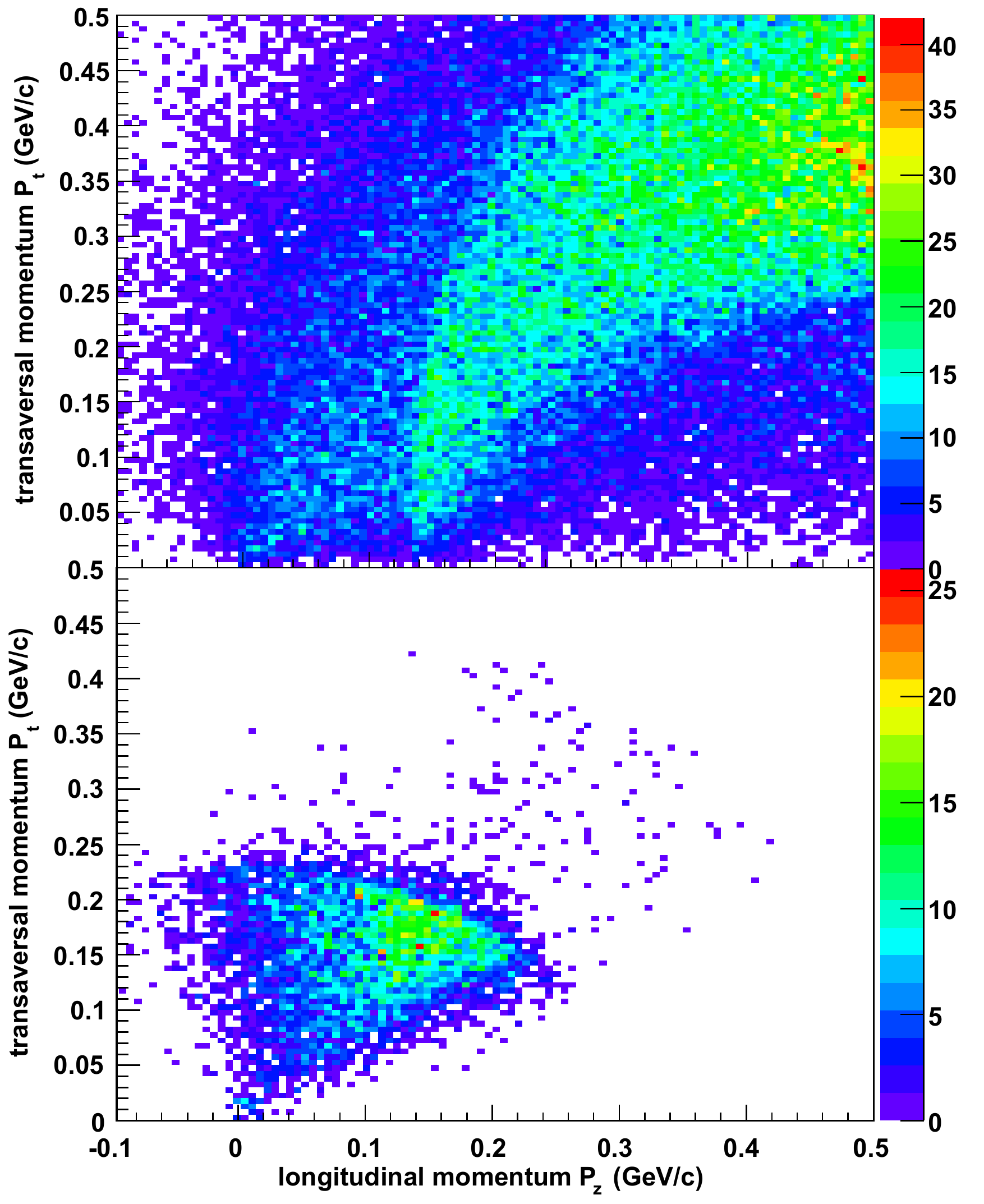}
    \caption{Distribution of transversal and longitudinal momenta smaller than 500\,MeV$/c$
			for $\Xi^-$ hyperons produced by $\mathrm{\overline{p}}$ interactions (top) and 
			for the fraction of hyperons stopped within the secondary target (bottom)~\cite{PhDLorente,PhysicsBook}.
			Hyperons with high momenta will dominantly decay before being slowed down.}
    \label{fig:XiStopPtPz}
  \end{center}
\end{figure}

\begin{figure}
  \begin{center}
    \includegraphics[width=\columnwidth]{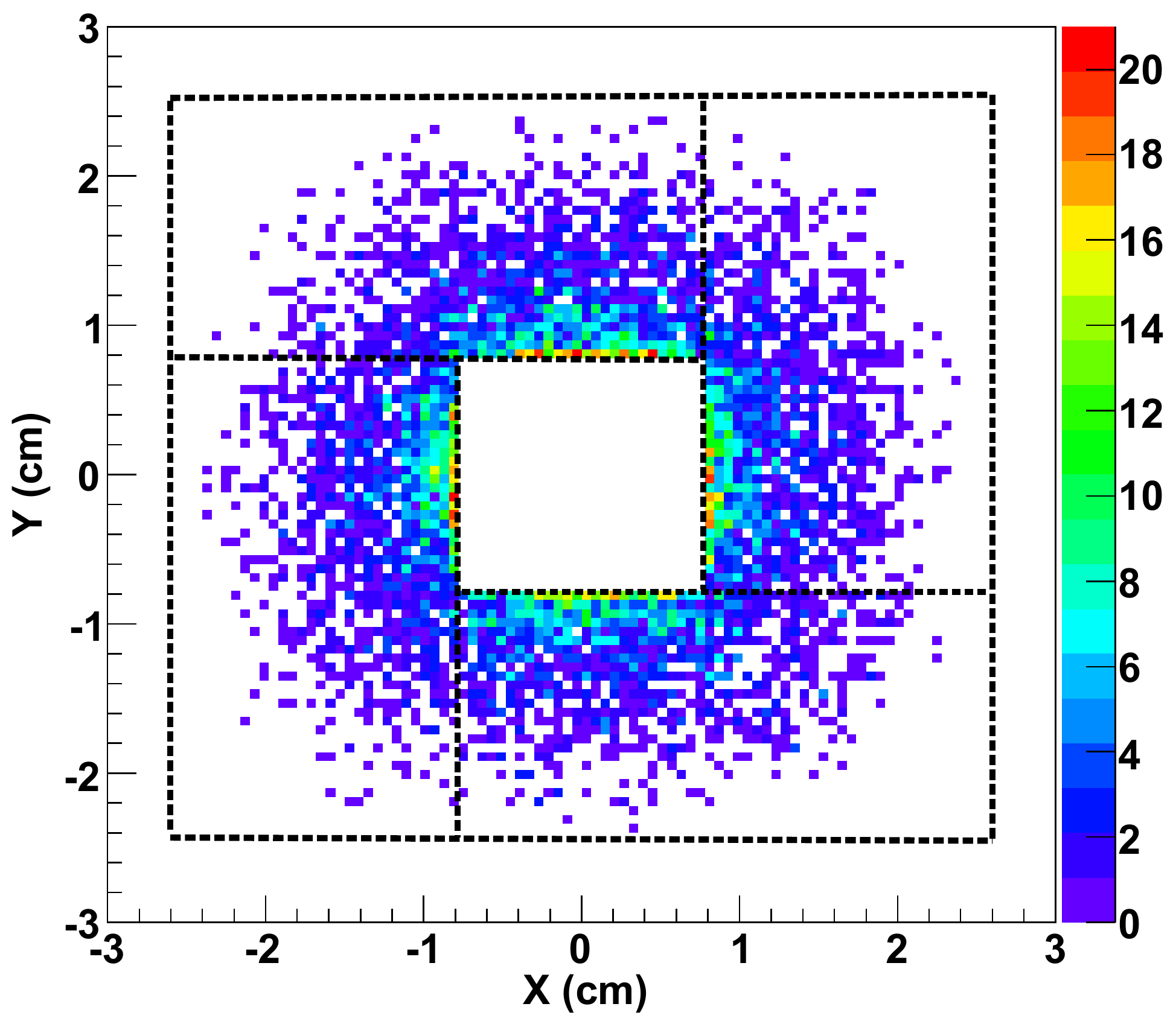}
    \caption{Distribution of stopping points of the $\Xi^-$ hyperons within
			the secondary target in the $X-Y$ plane transverse to the beam direction.
			The rectangles indicate the outline of the four active target
			segments~\cite{PhDLorente,PhysicsBook}. The volume of the active target 
			covers practically all stopping points.}
    \label{fig:XistopXY}
  \end{center}
\end{figure}

The main purpose of the hypernuclear target, seen in
Fig.~\ref{fig:SecondaryTarget}, is the tracking and stopping of the
produced cascade hyperons and their decay products.  The active part
of the secondary target was designed with silicon strip sensors of
dimensions 41 $\times$ 0.3 $\times$ 41\,mm$^3$ in a pitch of
150\,$\mu$m. 
The slowing down of the $\Xi^-$ proceeds (i) through a sequence of
nuclear elastic scattering events inside the residual nucleus in
which the annihilation has occurred and (ii) by energy loss
during the passage through an active absorber. If decelerated to
rest before decaying, the particle can be captured inside a
nucleus, eventually releasing two $\Lambda$ hyperons and forming
a double hypernuclei. The geometry of the target is essentially determined
by the lifetime of the hyperons and their stopping
time in solid material. 
Using an event
generator~\cite{Ferro2007} which is based on an Intra Nuclear
Cascade model and which takes as a main ingredient
the rescattering of the antihyperons and hyperons
in the target nucleus into account, the momentum distribution of outgoing
hyperons was calculated, Fig.~\ref{fig:XiStopPtPz}\,(top). Inside the hypernuclear 
target only hyperons with momenta
smaller than about 500 MeV$/c$ can be 
stopped prior to their free decay, Fig.~\ref{fig:XiStopPtPz}\,(bottom). 
The typical momenta of
the stopped hyperons are in the range of 200\,MeV$/c$. For the present
simulations a target thickness of 26\,mm was chosen
consisting out of 30 layers of silicon strip detectors
with alternating layers of absorber material, as shown in Fig.~\ref{fig:XistopXY}. The
active layers provide also tracking information on
the emitted weak decay products of the produced
hypernuclei. The choice of the absorber material is crucial for the magnitude
of the cross-section. Also the number of excited states of the
core should be small and the states should be well separated. 
The experiment will focus on light secondary
target nuclei with mass number $A_0 <$ 13.
Since the identification of the double hypernuclei
has to rely on the unique assignment of the detected
$\gamma$-transitions, different enriched
light isotopes ($^9$Be, $^{10,11}$B, $^{12,13}$C) will be used.

\section{High Purity Germanium Array}
\begin{figure*}
  \begin{center}
    \includegraphics[width=0.34\textwidth]{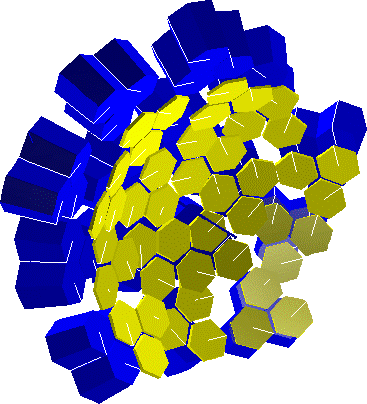}
    \includegraphics[width=0.55\textwidth]{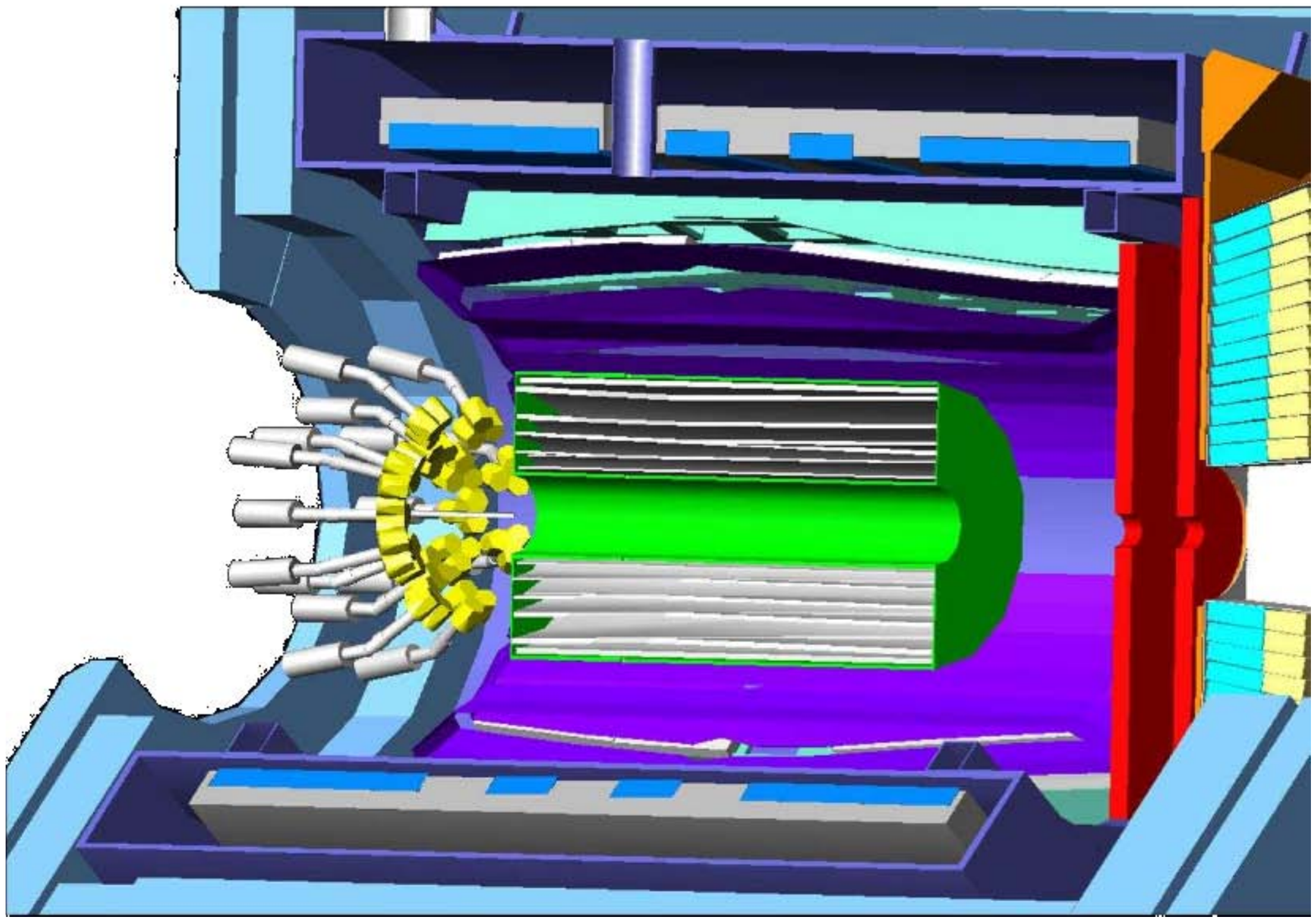}
    \caption{Arrangement of 15 cluster detectors  
      comprising a total of 45 HPGe crystals for
      hypernuclei experiments at \PANDA\ (left) and its 
      integration with electromechanical coolers into the
      target spectrometer when its end-cap and micro vertex 
      detectors have been removed (right)~\cite{PhDLorente}. 
      The beam enters from left; the primary and 
      secondary target are not visible.}
    \label{fig:HPGe}
  \end{center}
\end{figure*}

For the high resolution spectroscopy of excited hypernuclear states a
germanium $\gamma$-array is required. To
maximize the detection efficiency the $\gamma$-detectors must be
located as close as possible to the target. Hereby the main limitation
is the load of particles from background reactions. Most of the
produced charged particles are emitted into the forward region. Since
the $\gamma$-rays from the slowly moving hypernuclei are emitted
rather isotropically the germanium detectors will be arranged at
backward axial angles $\theta \geq$ 100$^\circ$.  A full simulation of
the hypernuclei detector's geometry has been
completed~\cite{PhDLorente}.  Fig.~\ref{fig:HPGe}\,(left) shows the design of the
$\gamma$-ray spectroscopy setup with 15 germanium cluster
detectors (each comprising 3 crystals). The integration of the
hypernuclear physics setup into the \PANDA\ target spectrometer is
shown in Fig.~\ref{fig:HPGe}\,(right).


At an average interaction rate of 5 $\cdot$ 10$^6$\,s$^{-1}$ UrQMD+SMM
calculations of $\bar{\mbox{p}} + $C interactions at 3\,GeV$/c$
momentum predict in the backward hemisphere a total charged and
neutral particle rate of 3.5 $\cdot$ 10$^7$\,s$^{-1}$. Since most of
the charged particles emitted into backward axial angles are very low
in kinetic energy, the majority of them will be absorbed in the beam pipe and
in the signal cables coming from the silicon sensors of the secondary
target. An especially 
critical issue are the neutrons emitted into backward axial
angles. 

Another major challenge at \PANDA\ is the operation of the
germanium detectors close to a strong magnetic field over long periods. 
These devices have been
only occasionally used in such conditions and their behavior
was not well known.
In order to look for a quantitative answer to these two
issues, an extensive R\&D project has been carried out within
a Joint Research Activity (HyperGamma) of the
European Union Sixth Framework Programme. Two
existing $\gamma$-ray detectors have been put inside a magnetic field
up to 1.6\,T: the Versatile and Efficient GAmma (VEGA)
super-segmented Clover detector and the Euroball Cluster
detector.
The experimental results obtained so far have been summarized
in Ref.~\cite{Lorente2007} and it was demonstrated that the energy resolution
of both detectors was nicely preserved up to 1 T.

\section{Particle Identification}
\begin{figure}
  \begin{center}
    \includegraphics[width=\columnwidth]{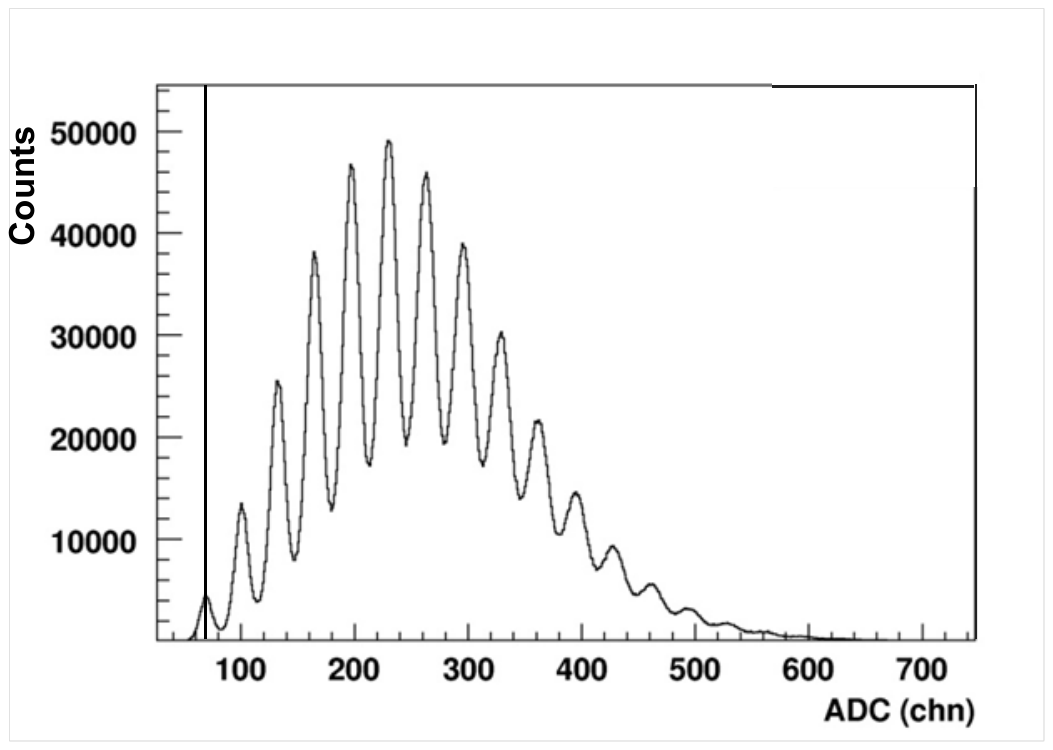}
    \caption{Measured pulse height spectrum from a SiPM$/$fiber
      setup, as being discussed as an option for the start
      detector of the \PANDA\ hypernuclear programme. The position
      of the pedestal peak is indicated by a vertical line, the
      following peaks resolve the signals from single and multiple
      pixels of the SiPM. On average, $\sim$ 5 pixels have fired in
      response to an electron crossing a thin organic fiber.}
    \label{fig:SiPMspectrum}
  \end{center}
\end{figure}

A small fiber barrel read-out by silicon photomultiplier (SiPM) has
been discussed as an option for a time-of-flight (TOF) start detector in
the hypernuclear physics programme of \PANDA. 
The SiPM is a novel semiconductor photodetector operated in
the limited Geiger mode, capable of resolving individual
photons~\cite{Dolgoshein2006}.  
 For this sub-detector system the achievable time resolution
at minimum detector mass is a main issue.  
A SiPM is intrinsically a very
fast detector with a single photoelectron time resolution of 
$<$ 100\,ps (FWHM). When coupled to thin and short scintillating fibers
the timing properties are fully dominated by the scintillation time
constants and depend only on the average number of detected photons.
The scintillation light from a 2\,m organic fiber with double cladding has been 
measured by two SiPM when excited by minimum ionizing
particles crossing its center. The photoelectron yield was derived 
from the ADC spectrum as shown in Fig.~\ref{fig:SiPMspectrum}.
On average, $\sim$5 pixels have fired in response to an electron crossing the fiber.
The time resolution was determined by taking
the time difference between the left and right signal simultaneously and was
found to be FWHM= 1.4\,ns. 
A gate on individual peaks made it
possible to determine the time resolution as a function of the number of
fired pixels. This study could be used 
to estimate possible improvements by increased light output.

A GEANT4 simulation was performed of such a TOF system: a start detector of $\sim$ 2000
scintillating fibers placed in two rings and a TOF barrel detector of
16 slabs (3 $\times$ 0.5 $\times$ 180\,cm). The simulation revealed
that for low
momentum kaon identification the stop detector must provide a time resolution of $<$ 100\,ps,
whereas the fiber detector has to provide the start time with a
minimum resolution of $\sim$ 400\,ps. It seems challenging to achieve
these values with the geometries and photon detection 
devices described so far. 

One improvement is made
by using a large area SiPM$/$fiber combination, and it was concluded
that SiPM with 3 $\times$ 3\,mm$^2$ 
active area can be used in combination with
scintillator or radiator strips of this cross-section 
for fast time response measurements. Only recently, 
lutetium aluminum garnet (LuAG, chemical formula Al$_5$Lu$_3$O$_{12}$) activated by cerium 
became available as possible fiber material.
Its density of 6.73\,g$/$cm$^3$ and decay time of 70\,ns brings some advantages for time 
dependent and coincidence measurements. Primarily, the higher density 
compared to organic fibers results
in a higher light output from the fibers.
The wavelength of the emission spectrum is at 535\,nm which is ideal for avalanche diode or SiPM readout. The photon yield is $\sim$ 20\,000 Photons$/$MeV and crystals can be grown 1\,mm in diameter and up to 500\,mm length. 
We have started the research into prototypes such as fiber bundles that could be used as a TOF start detector.

\section{Conclusions}
In combination with the high
luminosity of the anti-proton beam in HESR the \PANDA\ experiment at FAIR  will be able to explore 
the level scheme of double hypernuclei for the first time.
The spectroscopic information on double hypernuclei will be obtained via 
$\gamma$-ray detection using HPGe detectors located near a dedicated
arrangement of targets.

\IEEEtriggeratref{2}
\bibliographystyle{IEEEtran}
\bibliography{KAOS-28-07-08}

\end{document}